\documentclass[12pt,thmsa]{article}
\usepackage{sw20lart}
\usepackage{url}
\usepackage{amsmath}

\begin{document}

\author{Emilio Santos \and Departamento de F\'{i}sica. Universidad de Cantabria \and %
email address: \url{santose@unican.es} }
\title{Quantum field theory in the Weyl-Wigner representation.}
\date{September 2025}
\maketitle

\begin{abstract}
The Wigner representation for quantum mechanics of particles is generalized
to Bose fields. The standard Hilbert space quantization becomes, via the
Weyl transform, a quantization method that consists of adding a Gaussian
zeropoint field distribution to the vacuum. I comment on the possible
advantages of the method in order to study quantum fields in curved spaces.
I study a unified formulation of non-relativistic quantum electrodynamics in
the Weyl-Wigner formalism, in terms of (classical-like) c-numbers.
\end{abstract}

\tableofcontents

\section{Introduction}

The aim of the article is to generalize the Wigner representation, initially
proposed for quantum mechanics, to Bose fields. Thus I review and extend
previous work \cite{Universe}, \cite{FOOP}. The formalism provides an
alternative to the canonical field quantization, in terms of operators on a
Hilbert space, that might have advantages for the study of quantum fields in
curved spacetimes and for the search of a satisfactory quantum gravity
theory. In the last section of the article I present a formulation of
non-relativistic quantum electrodynamics involving particles and
electromagnetic field in a unified treatment. In order that the reader is
aware of the background, main developments, motivations and relevance of the
study it is appropriate to put the paper in a wider context as follows.

The article is a contribution to a research program attempting to elaborate
a realistic interpretation of quantum theory \cite{book}. This fits in the
aim of theoretical physics as it was well summarized in the initial
paragraph of the celebrated EPR article \cite{EPR}: ``Any serious
consideration of a physical theory must take into account the distinction
between the objective reality, which is independent of any theory, and the
physical concepts with which the theory operates. These concepts are
intended to correspond with the objective reality, and by means of these
concepts \textit{we picture this reality to ourselves}''\emph{\ }\textit{(}%
my emphasis\textit{)''}. I will name ``realistic'' those interpretations of
a physical theory which provide a picture of reality, although the name
might be disputed because ``realism'' is too broad a concept. The choice is
justified because in recent times the denomination ``local realism'' has
become popular in relation with Bell's work \cite{Bell}, see e.g. \cite
{EPJP25}.

The distinction between reality and theoretical concepts by EPR \cite{EPR}
means that different theories may exist for the description of a given
domain of reality. Therefore a specific theory may be very efficient in
order to derive predictions for the results of experiments but other
theories, or different formulations of the same theory, may be more suitable
to get a picture of reality. This fact is specially relevant for quantum
theory, which has had a large variety of interpretations \cite{Handbook}. I
believe that getting a picture of reality is an essential aspect of physics,
or at least an essential ingredient of the philosophical view about the
natural world. In fact a theory should deal with what nature does and not
only with the results of human observations or measurements. Hence a purely
pragmatic approach to physics is not sufficient for our understanding of the
world, say an approach which only appreciates the prediction of empirical
results, as is the Copenhagen interpretation. In my view this is not just a
matter of taste because the picture of reality provided by a physical theory
may be a clue for future progress.

Quantum field theory (QFT) with canonical quantization has had a spectacular
success in the quantitative prediction of observable properties, specially
in the domain of quantum electrodynamics. However there are difficulties in
other domains like quantum gravity or curved spaces. The former success show
that canonical QFT provides an extremely efficient algoritm for the
calculations, in particular renormalization techniques allow the removal of
divergences, but the subtraction of infinities is not satisfactory from a
fundamental point of view. I believe that the difficulty to extend QFT to
gravity or curved spaces has to do with the fact that neither canonical
quantization nor path integrals \cite{PATHS} provide pictures of reality.
Thus the aim of the present article is \textit{to show} that the Wigner
representation may be generalized to Bose fields and \textit{to suggest }%
that further work along the same line of research might solve the present
problems.

The Wigner representation is known from the early period of quantum
mechanics. The problem is that it was applied to non-relativistic quantum
mechanics, where it does not provide a ``picture of reality'', as commented
at the end of the next section. Indeed the success of the standard theory of
fundamental partcles show, or strongly suggets, that the inhabitants of our
world are quantum fields, while quantum particles consist of many
interacting fields, which is expressed saying that physical particles are
dressed by fields. The situation strongly contrasts with classical
mechanics, where point particles are usually good representations of bodies.
In my view the attempt to interpret quantum theory starting with
non-relativistic quantum mechanics has been a persistent error. I believe
that we should start with the interpretation of fields.

In summary I think that the problems of QFT might be solved or alleviated
with a method of quantization different from the standard ones (either
canonical of path integrals), and I propose the Wigner representation
toghether with a yet unknown extension to Fermi fields appropriate for a
realistic interpretation, which I hope is possible.

\section{The Wigner representation in quantum mechanics}

In the following I revisit the Wigner representation (or Weyl-Wigner, WW) of
quantum mechanics for non-relativistic particles. It is well known \cite
{Scully}, \cite{Zachos}, but I shall summarize the most relevant features
for convenience in later sections.

\subsection{Weyl transform and Wigner function}

In 1927 Weyl proposed a quantization method for systems of particles via a
transform that converts classical (c-number) coordinates and momenta into
operators in a Hilbert space \cite{Weyl} as follows. Let us consider a
system of N particles whose state is characterized by their position
coordinates, $x_{j},$ and momenta, $p_{j},$ j=1,2,...3N. Weyl transform
converts any polynomial function $f\left( \left\{ x_{j},p_{j}\right\}
\right) $ into a function of the quantum operators $\left\{ \hat{x}_{j},\hat{%
p}_{j}\right\} .$ (For the sake of clarity I will write quantum operators
with a `hat', e.g. $\hat{x}_{j},\hat{p}_{j}$ $,$ and numerical, c-number,
quantities without `hat', e.g. $x_{j},p_{j}).$ In one dimension Weyl
transform $T_{W}$ may be written

\begin{eqnarray}
&&T_{W}\left[ f\left( \left\{ x,p\right\} \right) \right] \equiv f_{W}\left(
\left\{ x,p\right\} \right)  \nonumber \\
&=&\frac{1}{4\pi ^{2}}\int d\lambda \int d\mu \int dx\int dp\exp \left[
i\lambda \cdot \left( x-\hat{x}\right) +i\mu \cdot \left( p-\hat{p}\right)
\right] f(x,p).  \label{00}
\end{eqnarray}
The generalization to 3N coordinates and momenta of N particles is
straightforward.

An important property of Weyl transform is that the function $f_{W}$
consists of operators in symmetrical order, this meaning a sum of terms with
one possible ordering each, divided by the number of terms. For instance 
\begin{equation}
T_{W}\left( x^{2}p\right) =\left( \hat{x}^{2}\hat{p}\right) _{sym}\equiv 
\frac{1}{3}\left( \hat{x}^{2}\hat{p}+\hat{x}\hat{p}\hat{x}+\hat{p}\hat{x}%
^{2}\right) .  \label{I5}
\end{equation}

Weyl transform is reversible (with caution, see below), and we are mainly
interested in the inverse of Weyl transform in this article. It leads from
operators, representing either states or observables, to functions in the
phase space of the particles, that is functions of their positions and
momenta. In one dimension the inverse Weyl transform is as follows 
\begin{eqnarray}
W_{\hat{M}}\left( x,p\right) &=&T_{W}^{rev}\left[ \hat{M}\right] \equiv 
\frac{1}{4\pi ^{2}}\int d\lambda \int d\mu \exp \left[ -i\lambda \cdot
x-i\mu \cdot p\right]  \nonumber \\
&&\times Tr\left\{ \hat{M}\exp \left[ i\lambda \cdot (x-\hat{x})+i\mu \cdot
\left( p-\hat{p}\right) \right] \right\} ,\smallskip \smallskip  \label{1}
\end{eqnarray}
where $\hat{M}$ is an operator for a quantum particle and $Tr\left\{
{}\right\} $ means the trace operation. It is possible to show that the
application of the Weyl transform followed by its inverse, or viceversa,
reproduces the initial expression, with the caution that follows.

Actually the definition eq.$\left( \ref{1}\right) $ is ambiguous and may
give rise to contradictions as shown with the following example. Let us
consider the inverse Weyl transform eq.$\left( \ref{1}\right) $ of the
operator products $\hat{x}\hat{p}$ and $\hat{p}\hat{x}$. In both cases the
result is $xp=px$. Then the application of that transform to both sides of
the fundamental quantum commutation rule 
\begin{equation}
\hat{x}\hat{p}-\hat{p}\hat{x}=i
\rlap{\protect\rule[1.1ex]{.325em}{.1ex}}h%
,  \label{commutator}
\end{equation}
would lead to the absurd result 
\[
xp-px=0=i
\rlap{\protect\rule[1.1ex]{.325em}{.1ex}}h%
. 
\]
The difficulty may be solved taking into account that the Weyl transform
leads always to symmetrized expressions. Therefore we must specify that the
inverse Weyl transform eq.$\left( \ref{1}\right) $ has a sense only if
applied to products of operators in symmetrical order, but this constraint
is not too restrictive. In fact any operator which may be written as a
product of the fundamental operators $\left\{ \hat{x}_{j},\hat{p}%
_{j}\right\} $ may be rewritten as a sum of products in symetrical order by
repeated application of the canonical commutation rule, eq.$\left( \ref
{commutator}\right) .$

In 1932 Wigner \cite{Wigner} proposed a representation of non-relativistic
quantum mechanics in terms of a formalism with classical flavour. He defined
a function $W_{\psi }\left( \left\{ x_{j},p_{j}\right\} \mathbf{,}t\right) $
in the phase space of a set of particles from the quantum wavefunction $\psi
\left( \left\{ x_{j}\right\} \mathbf{,}t\right) $ of the state as follows
(in one dimension) 
\begin{equation}
W_{\psi }\left( x,p\right) =\int du\psi \left( x-u\right) \psi ^{*}\left(
x+u\right) \exp \left( 2iup\right) ,  \label{01}
\end{equation}
where $W_{\psi }\left( x,p\right) $ is named Wigner function of the state $%
\psi $. The Wigner eq.$\left( \ref{01}\right) ,$ defined in the
Schr\"{o}dinger representation (that is, involving wave functions), is
equivalent to the inverse Weyl transform, eq.$\left( \ref{1}\right) ,$
formulated in the abstract Hilbert space formalism. Weyl\'{}s is more
general because it applies to any (trace-class) operator, for instanse
observables, in addition to states. Of course the Wigner eq.$\left( \ref{01}%
\right) $ may, and it has been, generalized to operators representing
observables.

\subsection{Equivalence with quantum mechanics}

Several properties of the Wigner representation are reported in the
following. Most relevant is the fact that the expectation value of the
observable $\hat{M}$ in the state $\psi $ becomes the integral 
\begin{equation}
\left\langle \psi \left| \hat{M}\right| \psi \right\rangle =\int W_{M}\left(
x,p\right) W_{\psi }\left( x,p,t\right) dxdp.  \label{2}
\end{equation}
Hence the expectation values obtained via the Wigner function agree with
those obtained from the canonical, Hilbert space, formalism of
non-relativistic quantum mechanics. This fact guarantees that the
predictions of the results of experiments should agree in both, canonical
and Wigner representations, provided that the laws of evolution are
appropriated.

The evolution of the Wigner function of a state may be obtained from the
time derivative of $W_{\psi }\left( r,p\mathbf{,}t\right) $ eq.$\left( \ref
{01}\right) $, taking Schr\"{o}dinger equation into account in order to
express $\partial \psi \left( r-u\mathbf{,}t\right) /\partial t$ and $%
\partial \psi ^{*}\left( r+u\mathbf{,}t\right) /\partial t$ as functions of $%
r-u$ and $r+u$ respectively. This should be followed by an integral with
respect to $u\mathbf{,}$ but the integration requires an explicit knowledge
of the dependence of $\psi \left( x,t\right) $ on $x$. For the particular
case of a single particle moving in a potential $V\left( \mathbf{r}\right) $
the evolution equation is as follows (in 3 dimensions)

\begin{eqnarray}
\frac{\partial W\left( \mathbf{r,p}\right) }{\partial t} &=&-\frac{1}{m}%
\mathbf{p}\cdot \mathbf{\nabla }W-\frac{1}{
\rlap{\protect\rule[1.1ex]{.325em}{.1ex}}h%
}\int \frac{d\mathbf{p}^{\prime }}{(2\pi )^{3}}\widetilde{V}\left( \mathbf{%
r,p}^{\prime }\right) W\left( \mathbf{r,p+p}^{\prime }\mathbf{,}t\right) , 
\nonumber \\
\widetilde{V}\left( \mathbf{r,p}^{\prime }\right) &\equiv &\int d\mathbf{u}%
\sin \left( \mathbf{p}^{\prime }\mathbf{\cdot u}\right) \left[ V\left( 
\mathbf{r+}
\rlap{\protect\rule[1.1ex]{.325em}{.1ex}}h%
\mathbf{u/}2\right) -V\left( \mathbf{r-}
\rlap{\protect\rule[1.1ex]{.325em}{.1ex}}h%
\mathbf{u/}2\right) \right] ,  \label{evo}
\end{eqnarray}
which is clearly more involved than Schr\"{o}dinger equation. A general
evolution equation of the Wigner function may be written in terms of the
coordinates $\left\{ x_{j}\right\} $ and momenta $\left\{ p_{j}\right\} $ of
several particles as follows

\begin{eqnarray}
\frac{\partial W}{\partial t} &=&\frac{2}{
\rlap{\protect\rule[1.1ex]{.325em}{.1ex}}h%
}\sin \left[ \frac{
\rlap{\protect\rule[1.1ex]{.325em}{.1ex}}h%
}{2}\left( \frac{\partial }{\partial x_{j}}\frac{\partial }{\partial
p_{j}^{\prime }}-\frac{\partial }{\partial p_{j}}\frac{\partial }{\partial
x_{j}^{\prime }}\right) \right] \left[ W\left( \left\{ x_{j}\mathbf{,}%
p_{j}\right\} \right) H_{part}\left( \left\{ x_{j}^{\prime }\mathbf{,}%
p_{j}^{\prime }\right\} \right) \right]  \nonumber \\
{} &\equiv &\frac{2}{
\rlap{\protect\rule[1.1ex]{.325em}{.1ex}}h%
}\sum_{n=0}^{3N}\frac{\left( -1\right) ^{n}}{\left( 2n+1\right) !}\left[ 
\frac{
\rlap{\protect\rule[1.1ex]{.325em}{.1ex}}h%
}{2}\left( \frac{\partial }{\partial x_{j}}\frac{\partial }{\partial
p_{j}^{\prime }}-\frac{\partial }{\partial p_{j}}\frac{\partial }{\partial
x_{j}^{\prime }}\right) \right] ^{2n+1}  \nonumber \\
&&\times \left[ W\left( \left\{ x_{j}\mathbf{,}p_{j}\right\} \right)
H_{part}\left( \left\{ x_{j}^{\prime }\mathbf{,}p_{j}^{\prime }\right\}
\right) \right] \equiv \left\{ W,H_{part}\right\} _{M}\text{ ,}
\label{Moyal}
\end{eqnarray}
where we should identify $\left\{ x_{j}^{\prime },p_{j}^{\prime }\right\}
=\left\{ x_{j},p_{j}\right\} $ after performing the derivatives. $\left\{
W,H_{part}\right\} _{M}$ is a simplified notation to be used in the
following, the subindex $M$ standing for \textit{Moyal bracket}. In eq.$%
\left( \ref{Moyal}\right) $ $\sin \left( x\right) $ stands for its expansion
in powers of $x$.

A remarkable property of eq.$\left( \ref{Moyal}\right) $ is that when either
the Hamiltonian or the Wigner function or both are at most quadratic in the
coordinates and momenta, only first and second derivatives may appear in eq.$%
\left( \ref{Moyal}\right) ,$ whence the Moyal bracket becomes the Poisson
bracket of classical dynamics. The evolutions of coordinates and momenta are
governed by classical laws in the Wigner representation in that case.
Typical example is a set of linearly coupled harmonic oscillators, for
instance the motion of ions in a lattice crystal using a standard
approximation.

The mentioned properties give rise the Wigner (or Weyl-Wigner, WW)
representation of non-relativistic quantum mechanics, which is equivalent to
the approach in terms of either operators or wavefunctions for the states.
In fact eq.$\left( \ref{2}\right) $ and the derivation of eqs.$\left( \ref
{evo}\right) $ and $\left( \ref{Moyal}\right) $ from Schr\"{o}dinger
equation guarantee that the expectation values obtained via the Wigner
representation agree with those got from the canonical formalism of quantum
mechanics, thus proving the equivalence of the formulations.

The Wigner representation is helpful for some purposes, but the calculations
are usually more involved than the standard ones, e.g. those using
Schr\"{o}dinger equation. Also it does not provide an intuitive
(``realistic'') interpretation of quantum mechanics because the Wigner
functions of states are positive definite but for a slight fraction of
quantum states, that is when the wavefunction is Gaussian \cite{Soto}.
Therefore the states cannot be interpreted as probability distributions in
phase space. In sharp contrast an intuitive interpretation of the
Weyl-Wigner representation for Bose fields is possible, as commented on the
next section. The reason for the difference between elementary quantum
mechanics and quantized electromagnetic field has been discussed elsewhere 
\cite{Universe}, \cite{FOOP}.

\section{Bose fields in the Weyl-Wigner representation}

\subsection{The massive neutral spin zero field}

\subsubsection{Classical treatment}

Let us start with a field fulfilling the Klein-Gordon equation, that is 
\begin{equation}
\left( -\frac{\partial ^{2}}{\partial t^{2}}+\triangle \right) \psi \left( 
\mathbf{r,}t\right) =m^{2}\psi \left( \mathbf{r,}t\right) ,  \label{2.1}
\end{equation}
where $\triangle $ is the Laplacean operator (in units $c= 
\rlap{\protect\rule[1.1ex]{.325em}{.1ex}}h%
=1,$ but Planck constant $
\rlap{\protect\rule[1.1ex]{.325em}{.1ex}}h%
$ will be restored for clarity in some cases). Eq.$\left( \ref{2.1}\right) $
appeared in the context of quantum mechanics, but it may be treated with the
methods of classical field theory. It was studied for the first time during
the pioneer work of Schr\"{o}dinger in 1925-26. Thus it may be appropriately
named ``relativistic Schr\"{o}dinger equation''. As is well known eq.$\left( 
\ref{2.1}\right) $ does not predict correctly the fine details of atomic
spectra, whence Schr\"{o}dinger himself restricted attention to the
non-relativistic approximation, which is an appropriate evolution equation
for non-relativistic quantum mechanics.

The classical treatment of eq.$\left( \ref{2.1}\right) $ may start with the
Lagrangean density 
\begin{equation}
L=\partial _{\mu }\psi ^{*}\cdot \partial ^{\mu }\psi -m^{2}\psi ^{*}\psi ,
\label{2.5}
\end{equation}
where $\partial _{\mu }$ are the derivatives with respect to the space-time
coordinates. Indeed Lagrange equations lead from eq.$\left( \ref{2.5}\right) 
$ to eq.$\left( \ref{2.1}\right) .$ From eq.$\left( \ref{2.5}\right) $ it is
also possible to get the energy-momentum tensor, that is 
\[
T_{\mu }^{\nu }=\sum \partial _{\mu }q\frac{\partial L}{\partial \left(
\partial _{\mu }q\right) }-L\delta _{\mu }^{\nu }. 
\]
In particular the Hamiltonian density may be written 
\begin{equation}
T_{00}=\frac{\partial \psi ^{*}}{\partial t}\frac{\partial \psi }{\partial t}%
+\nabla \psi ^{*}\nabla \psi +m^{2}\psi ^{*}\psi .  \label{2.6}
\end{equation}

In order to get the canonical (``second'') quantization of the field eq.$%
\left( \ref{2.1}\right) $ we shall start expanding the field in plane waves.
Here I assume that the field $\psi $ is real, which gives 
\begin{equation}
\psi =\sum_{j}\frac{1}{\sqrt{2\varepsilon _{j}}}\left[ a_{j}\exp \left( i%
\mathbf{k}_{j}\mathbf{.r-}i\varepsilon _{j}t\right) +a_{j}^{*}\exp \left( -i%
\mathbf{k}_{j}\mathbf{.r+}i\varepsilon _{j}t\right) \right] ,  \label{2.2}
\end{equation}
$a_{j}^{*}$ being the complex conjugate of $a_{j},$ where the single-mode
energy is 
\begin{equation}
\varepsilon _{j}=\sqrt{m^{2}+\mathbf{p}_{j}^{2}},\mathbf{p}_{j}= 
\rlap{\protect\rule[1.1ex]{.325em}{.1ex}}h%
\mathbf{k}_{j},  \label{Ej}
\end{equation}
and $\mathbf{p}_{j}$ is the linear momentum. In terms of the amplitudes $%
\left\{ a_{j},a_{j}^{*}\right\} $ the free field Hamiltonian may be got from
eq.$\left( \ref{2.6}\right) $ leading to 
\begin{equation}
H=\int d^{3}xT_{00}=\sum_{j}\varepsilon
_{j}a_{j}a_{j}^{*}=\sum_{j}\varepsilon _{j}\left| a_{j}\right| ^{2}.
\label{2.7}
\end{equation}

It is interesting to define the current density $j^{\mu }$ 
\begin{equation}
j^{\mu }=i\left( \psi ^{*}\partial _{\mu }\psi -\left( \partial _{\mu }\psi
^{*}\right) \psi \right) \Rightarrow \partial _{\mu }j^{\mu }=0,
\label{current}
\end{equation}
whence we may get a conserved quantity $Q$, that is

\begin{equation}
Q=\int j_{0}d^{3}x\text{ , }j_{0}=j^{0}=i\left( \psi ^{*}\frac{\partial \psi 
}{\partial t}-\frac{\partial \psi ^{*}}{\partial t}\psi \right) .  \label{Q}
\end{equation}
In the case of charged fields $Q$ may be interpreted as the electric charge.

\subsubsection{Canonical (Hilbert space) quantization}

The canonical method to quantize a field is to promote the quantities $%
\left\{ a_{j},a_{j}^{*}\right\} $ to be operators on a Hilbert space (HS),
fulfilling the commutation rules 
\begin{equation}
\left[ \hat{a}_{j},\hat{a}_{k}\right] =\left[ \hat{a}_{j}^{\dagger },\hat{a}%
_{k}^{^{\dagger }}\right] =0,\left[ \hat{a}_{j},\hat{a}_{k}^{\dagger
}\right] =\delta _{jk},  \label{com}
\end{equation}
$\delta _{jk}$\ being Kronecker delta. The operators $\hat{a}_{j}$ $\left( 
\hat{a}_{j}^{\dagger }\right) $ are usually named annihilation (creation)
operators of particles.

The commutation properties of the operators eqs.$\left( \ref{com}\right) $
may be related to the standard commutation rules of (non-relativistic)
quantum mechanics at equal times, that is 
\begin{equation}
\left[ \hat{x}_{j}\left( t\right) ,\hat{p}_{l}\left( t\right) \right] =i 
\rlap{\protect\rule[1.1ex]{.325em}{.1ex}}h%
\delta _{jl},  \label{9}
\end{equation}
where $i$ is the imaginary unit and $\delta _{jl}$ is Kronecker delta. In
fact introducing the following change of variables for the field amplitude $%
a_{j}$%
\begin{equation}
x_{j}\left( t\right) \equiv \frac{c}{\sqrt{2}\omega _{j}}\left( a_{j}\left(
t\right) +a_{j}^{*}\left( t\right) \right) ,p_{j}\left( t\right) \equiv 
\frac{i
\rlap{\protect\rule[1.1ex]{.325em}{.1ex}}h%
\omega _{j}}{\sqrt{2}c}\left( a_{j}\left( t\right) -a_{j}^{*}\left( t\right)
\right) ,  \label{10}
\end{equation}
it is possible to show that the free evolution of $a_{j}\left( t\right) $ is
given by the evolution of $\left[ x_{j}\left( t\right) ,p_{j}\left( t\right)
\right] $ as if they were the coordinate and the momentum of a classical
harmonic oscillator. Therefore the classical field may be treated formally
as a collection of oscillators. Consequently we may write the quantum
counterpart of eq.$\left( \ref{10}\right) $ as follows 
\begin{equation}
\hat{x}_{j}\left( t\right) \equiv \frac{c}{\sqrt{2}\omega _{j}}\left( \hat{a}%
_{j}\left( t\right) +\hat{a}_{j}^{\dagger }\left( t\right) \right) ,\hat{p}%
_{j}\left( t\right) \equiv \frac{i
\rlap{\protect\rule[1.1ex]{.325em}{.1ex}}h%
\omega _{j}}{\sqrt{2}c}\left( \hat{a}_{j}\left( t\right) -\hat{a}%
_{j}^{\dagger }\left( t\right) \right) .  \label{8}
\end{equation}
whence taking eq.$\left( \ref{9}\right) $ into account we get the
commutation rules for the field eq.$\left( \ref{com}\right) $.

The Hamiltonian for the free evolution in terms of the operators $\left\{ 
\hat{a}_{j},\hat{a}_{j}^{\dagger }\right\} $ is as follows 
\begin{equation}
\hat{H}=\frac{1}{2}\sum_{j}\varepsilon _{j}\left( \hat{a}_{j}\hat{a}%
_{j}^{\dagger }+\hat{a}_{j}^{\dagger }\hat{a}_{j}\right) ,  \label{H}
\end{equation}
and the evolution of the free quantum field from an inital state,
represented by the density operator $\hat{\rho},$ is given in the HS
formalism by the Heisenberg equation, that is 
\begin{equation}
\frac{d}{dt}\hat{\rho}=\frac{i}{
\rlap{\protect\rule[1.1ex]{.325em}{.1ex}}h%
}\left[ \hat{H},\hat{\rho}\right] .  \label{Heisenberg}
\end{equation}
For details see any book on quantum field theory. I use the notation of
Berestetskii et al. \cite{Lifshitz}.

\subsubsection{Weyl-Wigner quantization}

Up to here the standard (Hilbert space) quantum theory of the said Bose
field. The Weyl-Wigner (WW) representation of \textit{the same quantum field}
is achieved via the (inverse) Weyl transform, $T_{W},$ which becomes, for a
quantum field operator $\hat{M},$ 
\begin{eqnarray}
T_{W}\left[ \hat{M}\right] &=&\frac{1}{(2\pi ^{2})^{n}}\prod_{j=1}^{n}%
\int_{-\infty }^{\infty }d\lambda _{j}\int_{-\infty }^{\infty }d\mu _{j}\exp
\left[ -2i\lambda _{j}\operatorname{Re}a_{j}-2i\mu _{j}\operatorname{Im}a_{j}\right] 
\nonumber \\
&&\times Tr\left\{ \hat{M}\exp \left[ i\lambda _{j}\left( \hat{a}_{j}+\hat{a}%
_{j}^{\dagger }\right) +\mu _{j}\left( \hat{a}_{j}-\hat{a}_{j}^{\dagger
}\right) \right] \right\} \equiv W_{\hat{M}}.\smallskip \smallskip
\smallskip \smallskip \smallskip \smallskip \smallskip  \label{Wtrans}
\end{eqnarray}
where $T_{W}\left[ \hat{M}\right] $ stands for (inverse) Weyl transform of
the operator $\hat{M},$ which may be either an observable or the density
operator of a state. This generalizes the Weyl transform eq.$\left( \ref{1}%
\right) $ from quantum mechanics of particles to quantum fields, taking into
account the fact that the real and imaginary parts of the field amplitude
evolve like the position and momentum of a harmonic oscillator \cite
{Universe}, \cite{FOOP}.

The result $W_{\hat{M}}\left( \left\{ a_{j},a_{j}^{*}\right\} \right) $
obtained from eq.$\left( \ref{Wtrans}\right) $ is a function of (c-number)
field amplitudes. Thus we might question whether the effect of the inverse
Weyl transform is not just reversing the HS quantization of the field
(involved operators) to become again a classical field (involving
classical-like amplitudes). The answer to the question is negative because
the WW formalism disagrees from the classical one in a fundamental aspect,
that is the definition of the ground state. In fact the ground state of a
classical field is the (empty) vacuum where all amplitudes are nil, that is $%
a_{j}=a_{j}^{*}=0$ for every $j$. In contrast the ground state of the
Hilbert-space quantized field is represented by the operator 
\begin{equation}
\hat{\rho}=\mid 0\rangle \langle 0\mid ,  \label{3}
\end{equation}
fulfilling the following equalities 
\begin{equation}
\hat{a}_{j}\mid 0\rangle =0\Rightarrow \langle 0\mid \hat{a}_{j}^{\dagger
}=0,\text{ for all radiation modes,}  \label{4}
\end{equation}
$0$\ being here the null vector in the Hilbert space. Usually this state is
called ``vacuum'' state, but it is not empty in the WW formalism. In fact if
the operator eq.$\left( \ref{3}\right) $ is inserted in eq.$\left( \ref
{Wtrans}\right) $ in place of $\hat{M},$ after some algebra we get the
Wigner function of that ``vacuum'' state, that is 
\begin{equation}
W_{0}\left( \left\{ a_{j}\right\} \right) =\prod_{j}\frac{2}{\pi }\exp
\left( -2\left| a_{j}\right| ^{2}\right) ,  \label{5}
\end{equation}
where we have normalized it for the integration with respect to $\prod_{j}d%
\operatorname{Re}a_{j}d\operatorname{Im}a_{j}.$ Hence the mean square value of each
amplitude in the vacuum is as follows 
\begin{equation}
\left\langle \left| a_{j}\right| ^{2}\right\rangle =\frac{1}{2}.  \label{6}
\end{equation}

Eq.$\left( \ref{5}\right) $ strongly suggests an interpretation of the
vacuum Wigner function as a probability distribution, whence eq.$\left( \ref
{6}\right) $ would be the variance of $\left| a_{j}\right| .$ The field
represented by eq.$\left( \ref{5}\right) $ is labeled the ``zeropoint
field'' (ZPF) and corresponds to the ``quantum vacuum fluctuations'' of the
canonical (Hilbert space) formalism for quantum fields.

There are also differences between HS and WW for other states. In the HS
formalism, all pure excited states of the field may be obtained by repeated
application of the creation operators to the vacuum state, and additional
pure states may be got which are linear combinations of the former. The
Wigner functions of these states may be obtained in the WW formalism using
the Weyl transform eq.$\left( \ref{Wtrans}\right) .$ However those Wigner
functions cannot be interpreted as probability distributions because they
are not positive in general. This fact might led some authors to think that
either WW representation for fields does not admit a realistic
interpretation or it is not equivalent to the standard HS one.

In my opinion the dilemma derives from the flawed assumption that all
state-vectors declared states in the HS representation are physical. For
instance a single-particle state in the form of a plane wave extended over
an arbitrary large normalization volume cannot represent a \textit{physical}
state of a (localized) particle, I believe. Indeed in the WW representation
particles (bosons) do not appear explicitly. We see just continuous fields,
whence the particles of the HS representation may be just useful
mathematical elements that appear in intermediate stages of the
calculations. The question has been discussed elsewhere \cite{FOOP}, \cite
{Universe}, in relation with the electromagnetic field and similar arguments
apply for other Bose fields. Further comments on the question are made in
subsection 5.4 below.

The evolution of the field amplitudes is given by the WW counterpart of the
Heinsenberg eq.$\left( \ref{Heisenberg}\right) $ in the HS formalism, that
is Moyal eq.$\left( \ref{Moyal}\right) .$ However this should be written in
terms of the field amplitudes, rather than coordinates and momenta, see eq.$%
\left( \ref{10}\right) $. That is

\begin{eqnarray}
\frac{\partial W_{\hat{M}}}{\partial t} &=&2\{\sin \left[ \frac{1}{4}\left( 
\frac{\partial }{\partial \operatorname{Re}a_{j}^{\prime }}\frac{\partial }{%
\partial \operatorname{Im}a_{j}^{\prime \prime }}-\frac{\partial }{\partial 
\operatorname{Im}a_{j}^{\prime }}\frac{\partial }{\partial \operatorname{Re}%
a_{j}^{\prime \prime }}\right) \right] \smallskip  \nonumber \\
&&\times W_{\hat{M}}\left\{ a_{j}^{\prime },a_{j}^{*\prime },t\right\}
H\left( a_{j}^{\prime \prime },a_{j}^{*\prime \prime }\right)
\}_{a_{j}},\smallskip  \label{Moyala}
\end{eqnarray}
where $\left\{ {}\right\} _{a_{j}}$ means making $a_{j}^{\prime
}=a_{j}^{\prime \prime }=a_{j}$ and $a_{j}^{*\prime }=a_{j}^{*\prime \prime
}=a_{j}^{*}$ after performing the derivatives. A remarkable result is that
the free field Hamiltonian $\left\{ a_{j},a_{j}^{*}\right\} $ is a quadratic
function of the amplitudes, whence only terms with first and second
derivatives remain in eq.$\left( \ref{Moyala}\right) .$ The consequence is
that the Moyal bracket becomes the Poisson bracket of classical dynamics,
whence the WW quantized free EM field evolution is governed by classical
theory (for a similar feature in the quantum electromagnetic field see \cite
{Universe}, \cite{FOOP}).

In summary we have started with the classical field eqs.$\left( \ref{2.1}%
\right) $ and $\left( \ref{2.5}\right) .$ After that we have quantized the
field via the canonical Hilbert space method promoting amplitudes to be
creation and annihilation operators of particles. Then we have performed an
inverse Weyl transform and got a theory with classical flavour. Indeed it
contains (c-number) amplitudes and classical evolution, which might appear
as if we returned to the classical theory eqs.$\left( \ref{2.1}\right) $ and 
$\left( \ref{2.5}\right) $. However there is a relevant variation. At a
difference with a classical field, in the WW quantized field we assume the
existence of a non-empty vacuum filled with a random radiation eq.$\left( 
\ref{5}\right) .$ In conclusion I have achieved the goal of getting a
formalism for the quantized bose field that leads to a clear picture of
reality: it is just the classical field with the additional assumption that
the ground state (the ``vacuum'') consists of a random background radiation.
Particles (bosons) are mathematical elements, useful for computations, but
without physical reality.

\subsection{The charged spin-zero massive field}

The assumption that the field function $\psi $ in eq.$\left( \ref{2.1}%
\right) $ is real (in the mathematical sense of being defined in the real
numbers) leads to the quantum theory of \textit{neutral }spin zero massive
particles. Now we shall study the case when $\psi $ is complex. The
expansion in plane waves is similar to eq.$\left( \ref{2.2}\right) $ except
that now we shall introduce two different amplitudes, namely $a_{j}$ and $%
b_{j},$ obtaining the following

\begin{equation}
\psi \left( \mathbf{r,}t\right) =\sum_{j}\frac{1}{\sqrt{2\varepsilon _{j}}}%
\left[ a_{j}\exp \left( i\mathbf{k}_{j}\mathbf{.r-}i\varepsilon _{j}t\right)
+b_{j}^{*}\exp \left( -i\mathbf{k}_{j}\mathbf{.r+}i\varepsilon _{j}t\right)
\right] ,  \label{2.3}
\end{equation}
where $b_{j}^{*}$ \textit{is not} the complex conjugate of $a_{j}.$ The main
difference with the canonical quantization of the neutral field is that now
we must introduce two kinds of field operators namely $\left\{ \hat{a}_{j},%
\hat{a}_{j}^{\dagger }\right\} $ and $\left\{ \hat{b}_{j},\hat{b}%
_{j}^{\dagger }\right\} $ which correspond to different kind of particles.
These operators fulfil the following commutation rules 
\begin{equation}
\left[ \hat{a}_{j},\hat{a}_{k}^{\dagger }\right] =\delta _{jk},\left[ \hat{b}%
_{j},\hat{b}_{k}^{\dagger }\right] =\delta _{jk},  \label{2.4}
\end{equation}
all other commutators being nil. The operators $\hat{a}_{j}^{\dagger }$ and $%
\hat{b}_{j}^{\dagger }$ are interpreted as creating particles and
antiparticles, respectively, while $\hat{a}_{j}$ and $\hat{b}_{j}$
annihilate them. Particles and antiparticles have the same mass but, if they
are charged, opposite electric charges, see any book on quantum field theory 
\cite{Lifshitz}.

The quantization in the WW representation is quite similar to the case of
neutral particles discussed above. The Weyl transform eq.$\left( \ref{Wtrans}%
\right) $ now consists of the product of two similar terms involving the
particle\'{}s and antiparticle\'{}s operators respectively. Then the vacuum
consists of two fields with opposite charge and a Wigner function similar to
eq.$\left( \ref{00}\right) $ each. That is the Wigner function of the vacuum
state for the two fields should be 
\begin{equation}
W_{0}\left( \left\{ a_{j},b_{k}\right\} \right) =\prod_{j}\frac{2}{\pi }\exp
\left( -2\left| a_{j}\right| ^{2}\right) \times \prod_{k}\frac{2}{\pi }\exp
\left( -2\left| b_{k}\right| ^{2}\right) .  \label{11}
\end{equation}
If the fields are charged the positive and negative electric charges of the
vacuum may cancel on the average, they being associated to the amplitudes $%
a_{j}$ and $b_{k}$ respectively. However the vacuum field eq.$\left( \ref{11}%
\right) $ may give rise to fluctuations of both energy and charge in the
vacuum. Actually particles and antiparticles interact via the
electromagnetic force whence the actual vacuum should take the
electromagnetic interaction of the charges into account. Consequently the
vacuum field will be more involved than eq.$\left( \ref{11}\right) $ where
the electromagnetic interactions are neglected.

\subsection{Spin-one fields. Electromagnetism}

Massive spin-one fields, either neutral or charged, are studied via the
Proca equation. The Weyl-Wigner quantization is similar to that for
spin-zero fields and will not be studied further here \cite{Lifshitz}.

The most relevant spin-one field is the quantized electromagnetism, whose
Weyl-Wigner representation has been studied elsewhere \cite{Universe}, \cite
{FOOP}, and I will briefly revisit in the following. It shall be studied in
the Coulomb gauge, as is appropriate for the study of the interaction of the
field with particles in non-relativistic motion, which will be made in the
last section 5.

The free EM field Hamiltonians in the HS and WW formalisms are related by
the inverse Weyl transform and are defined as follows \cite{Universe} 
\begin{equation}
\hat{H}_{HS}=
\rlap{\protect\rule[1.1ex]{.325em}{.1ex}}h%
\sum_{j}\omega _{j}(\hat{a}_{j}^{\dagger }\hat{a}_{j}+\frac{1}{2}),H_{WW}= 
\rlap{\protect\rule[1.1ex]{.325em}{.1ex}}h%
\sum_{j}\omega _{j}\left| a_{j}\right| ^{2},  \label{Hamiltonians}
\end{equation}
respectively.

When there are charged particles the evolution of both the particles and the
field is modified by the interactions, the interaction Hamiltonian being
given in WW by 
\begin{equation}
H_{int}=-\sum_{k}e_{k}\mathbf{p}_{k}\mathbf{\cdot A}\left( \mathbf{x}%
_{k},\left\{ a_{j},a_{j}^{*}\right\} \right) ,  \label{3.20}
\end{equation}
where $e_{k}$ is the charge of the particle, $\mathbf{p}_{k}$ its momentum,
and $\mathbf{A}$ is the potential vector of the field at the position $%
\mathbf{x}_{k}$ of the particle. The subindex $k$ runs for all charged
particles, but we might sum for all particles in eq.$\left( \ref{3.20}%
\right) $ substituting $j$ for $k$ and putting $e_{j}=0$ for neutral
particles. I note that $\mathbf{p}_{k}$ is a 3D vector which corresponds to
3 generalized one-dimensional momenta. I omit the expression for the
dependence of the potential vector $\mathbf{A}$ in terms of the field
amplitudes $\left\{ a_{j},a_{j}^{*}\right\} ,$ which is well known. Actually
the work in the Coulomb gauge requires taking also into account the
instantaneous electrostatic interaction between charged particles, which is
not included in the interaction Hamiltonian eq.$\left( \ref{3.20}\right) .$
Here I skip this term but it should be included in actual calculations.

In the WW formalism the field evolution is given by the Moyal eq.$\left( \ref
{Moyala}\right) ,$ but it is governed by the classical, Maxwell, theory when
the total Hamiltonian is at most quadratic in the amplitudes, as is usually
the case. This is similarly to other Bose fields studied in section 3. The
most relevant difference between the canonical, HS, and the WW formalism of
the field is that the WW quantized field assumes the existence of a
radiation in the vacuum with a distribution of amplitudes eq.$\left( \ref{5}%
\right) .$ That is the zeropoint field (ZPF) which corresponds to the
``quantum vacuum fluctuations'' of the standard (Hilbert space) formalism
for the quantized electromagnetic field.

Te main result is that both formalisms for the quantized EM field are
equivalent, that is canonical (HS) and WW. Both would predict the same
results for experiments.The field $E_{ZPF}$ is random (mathematically it is
a stochastic process) whose most relevant property is the energy density, $%
\rho \left( \omega \right) ,$ defined as the energy per unit time, unit
volume and unit frequency interval. This may be derived from the
distribution of the amplitudes of the radiation modes eq.$\left( \ref{5}%
\right) ,$ which have a mean energy $\frac{1}{2} 
\rlap{\protect\rule[1.1ex]{.325em}{.1ex}}h%
\omega $ per normal mode \cite{Universe}. Hence the total energy density per
unit frequency interval, $\rho \left( \omega \right) $, is the product of $%
\frac{1}{2}
\rlap{\protect\rule[1.1ex]{.325em}{.1ex}}h%
\omega $ times the number density of modes per unit frequency interval in a
large normalization volume, which gives 
\begin{equation}
\rho \left( \omega \right) =\frac{1}{2\pi ^{2}c^{3}} 
\rlap{\protect\rule[1.1ex]{.325em}{.1ex}}h%
\omega ^{3}.  \label{ZPF}
\end{equation}
This expression for the vacuum energy density (usually named zeropoint
field, ZPF) goes back to the early period of quantum theory, when it was
studied by Planck, Einstein, Nernst and other people (see e.g. \cite{Milonni}%
, \cite{Boyer18}). Eq.$\left( \ref{ZPF}\right) $ implies that the electric
and magnetic fields of the ZPF are very strong for high frequencies. However
the fields fluctuate rapidly then and the clean effect on charges is
relatively small.

\section{Discussion and applications}

In the previous section I have presented a quantization procedure of Bose
fields, which is physically equivalent to the canonical one. The
quantization consists of adding to the classical field a Gaussian
distribution of field amplitudes which correspond to the quantum vacuum
fluctuations in the canonical formalism, but the evolution remains
classical. In this section I discuss the realistic interpretation, related
work and possible applications.

\subsection{The realistic interpretation}

In my view the most interesting feature of the WW formalism for Bose fields
is that it allows a \textit{realistic interpretation }in the sense of
section 1. Actually after one century of quantum mechanics there is no
consensus about the interpretation \cite{Handbook}, Furthermore there is a
widespread state of opinion that a realistic interpretation is not possible,
but I do not agree \cite{book}, \cite{EPJP25}. The realistic interpretation
of Bose fields quantized via the WW representation is as follows:
Quantization means assuming that the classical laws are valid, but also that
the vacuum is not empty, it is filled with \textit{real fields} having a
Gaussian probability distribution, eq.$\left( \ref{5}\right) .$ That field
corresponds to the quantum vacuum fluctuations of the canonical formulation.

As commented on subsection 3.1 in WW quantized Bose fields the bosons are
not physical (localized) particles, just mathematical elements useful for
calculations. This is in particular the case for photons. In fact many
quantum phenomena taken as proofs of the existence of photons (more properly
a discrete character of the electromagnetic field) may be understand in
terms of a continuous field, at least qualitatively. This is the case in
particular for the photoelectric or the Compton effects \cite{Foundations}.

\subsection{Stochastic electrodynamics}

The essential ingredient of WW quantization is the existence of a vacuum
field (the zero-point field, ZPF) with a Gaussian distribution of field
amplitudes, see eq.$\left( \ref{5}\right) .$ Historically the idea of ZPF
appeared in the second Planck radiation theory of 1912. As said above
Einstein and other people were interested in it. Later on Walter Nernst
suggested that the ZPF might be relevant in order to explain the stability
of atoms and molecules. For a good account of these works see the book by
Milonni \cite{Milonni}. The ZPF line of research was abandoned due to the
appearance of Bohr atomic model in 1913, which led the community to a
different research program named ``old quantum theory'', that culminated in
1925 with the standard form of quantum mechanics. However around 1960 the
early idea of Nernst reappeared as a theory named ``stochastic (or random)
electrodynamics'' (SED)\ \cite{Braffort}, \cite{Marshall}. Many articles
have been published on the subject from that time. For a review of the work
made before 1996 see \cite{Dice}, more recent reviews are \cite{Boyer18} and
Chapter 5 of the book \cite{book} reproduced in \cite{arxiv}

Initially the purpose of SED was to explain some quantum non-relativistic
phenomena as due to the ZPF. Indeed there is an analogy between SED and
non-relativistic QED \cite{EPJP}. Later on the aim has been extended to the
relativistic domain defining SED more generally as ``classical electron
theory with classical electromagnetic zero-point radiation'' \cite{Boyer13},
a definition that might be applied to the WW quantized field.

\subsection{The lack of a WW representation for Fermi fields}

The quantization formalism studied in this paper has the shortcoming that it
cannot be easily extended to Fermi fields. If the generalization was
achieved it would provide a new and interesting formulation of the whole
quantum field theory. However in the restricted non-relativistic domain
Fermi fields may be treated as sets of particles, specially those with
spin-1/2 like electrons. In addition in that domain electromagnetism is the
unique quantum field of interest. Hence it is possible to formulate the
whole non-relativistic quantum electrodynamics in the Weyl-Wigner
representation because both the EM field and the particles may be WW
quantized. The resulting theory, i.e. non-relativistic QED will be sketched
in the next section 5.

\subsection{The divergence problem}

A possible difficulty of the approach studied in this paper is the
divergence of the vacuum fields, a problem that actually appears also in the
canonical Hilbert space formulation of quantum fields, although less
explicitly. As is well known the difficulty has been treated in practice
with renormalization techniques that have achieved a spectacular success.
However from a fundamental point of view the problem remains. A practical
solution is to state that the present treatment of fields cannot be correct
beyond Planck\'{}s density, but this gives rise to a difficulty known as the
``cosmological constant problem''\cite{Weinberg}. I may especulate about the
complete solution of the divergence problem, rather than just a practical
one, suggesting two possibilities. The first one is that the vacuum ZPF of
Fermi fields might provide a negative energy that could balance the positive
divergent energy of Bose fields. Another possibility is that the set of all
fields and interactions provide energy and pressure in free space with the
``equation of state'' of the vacuum, that is a negative pressure $p$ and
positive energy $\varepsilon $ such that $p=-\varepsilon ,$ which might be
counterbalanced by a cosmological constant in Einstein equation of general
relativity. Anyway the cancelation might not be exact, the remaining energy
and pressure giving rise to the cosmological dark energy \cite{Entropy}.

\subsection{Fields in curved spacetime}

The study of quantum fields in curved spacetime presents a difficulty,
namely that the commutation relations at neighbour points are well defined
in flat (Minkowski) space but not so clearly in curved space. The standard
method to avoid the problem is to define normal modes of the quantum fields
in a curved spacetime \cite{Birrell}, \cite{Parker}, then studying field
excitations via the creation of virtual particles. However this gives rise
to the non-uniqueness of canonical field quantization \cite{Fulling}. I
believe that the WW quantization might avoid, or diminish, these
difficulties of the canonical quantization.

Several predictions have been achieved with the study of quantum fileds in
curved space, the most popular being the Hawking radiation by black holes.
Here I will comment on another one, the Unruh effect. It is currently
interpreted stating that photon detectors accelerating through the quantum
vacuum behave as if they were located in an inertial frame in a thermal bath
with temperature 
\begin{equation}
T=\frac{
\rlap{\protect\rule[1.1ex]{.325em}{.1ex}}h%
a}{2\pi ck_{B}},  \label{bath}
\end{equation}
$a$ being the acceleration and k$_{B}$ Boltzmann constant \cite{Davies}, 
\cite{Unruh}, \cite{Crispino}. On the other side Timothy Boyer has derived
eq.$\left( \ref{bath}\right) $ as due to the modification of the ZPF
spectrum of the vacuum, eq.$\left( \ref{ZPF}\right) ,$ when it is seen in an
accelerating frame, see \cite{Boyer13} and references therein. On the other
hand the ``detection of individual photons'' could be explained as the
action of a continuous field on a photocounter, say via a photoelectric
effect \cite{FOOP}, \cite{Foundations}. Boyer attributes the agreement to
the fact that ``classical electron theory with classical electromagnetic ZPF%
\textit{''} is close to the quantum theory of the EM field. In this article
I have shown that it is not just close, but identical. Indeed the sentence
within inverted commas is actually an appropriate definition for \textit{WW
quantized electromagnetic field}.

The two mentioned interpretations of the Unruh effect illustrate the
difference beween the canonical and the WW quantization. In the former
approach the current interpretation is that in curved spacetime the vacuum
is excited via the creation of particles (photons) with a thermal
distribution. In the latter the spectrum of the ZPF differs from eq.$\left( 
\ref{ZPF}\right) $ in an accelerated frame, that is the distribution of
energy amongst the frequencies of the radiation modes.

\subsection{Quantum gravity}

As is well known the gravitational field (general relativity) cannot be
quantized via the standard method of expanding the field in plane waves
because the fundamental (Einstein) equation is not linear. The nonlinear
character does not fit in the quantum commution rules. The Unruh effect has
been sometimes taken as a ``quantum gravity'' effect. Indeed it combines a
quantum element (the electromagnetic ZPF) with gravity (or acceleration).
However quantum gravity is mainly devoted to allow calculations in cases
when both quantum theory and gravity are relevant (typically involving mass
densities of order Planck\'{}s). A large amount of work has been made on the
problem \cite{Rovelli}. Both canonical and path integrals quantization have
been used without achieving a completely satisfactory quantum gravity theory
till now.

The gravitational field is bosonic and therefore I believe that it could be
quantized via the addition of a vacuum field to the classical theory. This
would amount to assuming the existence of Gaussian spacetime fluctuations at
short distances. I do not know how the WW quantization might help for the
formulation of a quantum gravity theory. In any case a problem would arise
due to the small value of the Newton constant, which leads to a huge scale
difference between gravitational and typical quantum field theory vacuum
energy.

\section{Non-relativistic quantum electrodynamics}

Relativistic quantum electrodynamics (QED) cannot be formulated within WW
because for Fermi fields a transform playing the role Weyl\'{}s for Bose
fields is not available. However it is possible to formulate
non-relativistic QED, WW-quantized, treating Fermi particles in the
non-relativistic approximation where the Wigner representation is well known 
\cite{Scully}, \cite{Zachos}, see the Introduction section. Furthermore it
is possible a unified treatment of both the particles and the
electromagnetic (EM) field, in terms of generalized coordinates $x_{j}$ and
momenta $p_{j}$ as shown in the following. That theory has purely academic
interest because the problem has been studied using most appropriate
relativistic quantum electrodynamics (QED). The interest of the
non-relativistic theory might be to illustrate the calculational methods of
the Weyl-Wigner (WW) representation in comparison with either canonical (HS)
or path integrals treatments.

\subsection{Weyl-Wigner unified treatment of quantum particles and EM field}

Quantum mechanics of particles in the Weyl-Wigner (WW) representation was
skecthed in section 2 and the WW \textit{quantized} EM field has been
studied in subsection 3.3 and elsewhere \cite{Universe}, \cite{FOOP}. Now
let us consider a system of $N$ quantum particles, some of them charged and
therefore interacting with the EM field.

The typical problem of non-relativistic QED in the Weyl-Wigner formalism is
the joint evolution of the state of the particles represented by the Wigner
function in terms of the coordinates and momenta $\left\{
x_{j},p_{j}\right\} ,$ and the fields represented also by the Wigner
function in terms of the amplitudes $\left\{ a_{l},a_{l}^{*}\right\} .$ We
need the total Hamiltonian, that is 
\begin{equation}
H_{tot}=H_{part}\left( \left\{ x_{j},p_{j}\right\} \right) +H_{field}\left(
\left\{ a_{l},a_{l}^{*}\right\} \right) +H_{int}\left( \left\{
x_{j},p_{j},a_{l},a_{l}^{*}\right\} \right)  \label{7}
\end{equation}
The evolution of the system would follow from the application of Moyal
equations, that is eq.$\left( \ref{Moyal}\right) $ for the particles and eq.$%
\left( \ref{Moyala}\right) $ for the field, but the combination of the
particles coordinates and momenta with the field amplitudes is inconvenient
in practice.

A more refined method would be to use for the field (pseudo) coordinates $%
x_{j}$ and momenta $p_{j}$ of the plane waves expansion, using the change of
variables eq.$\left( \ref{10}\right) .$ Then we might apply a unique Moyal
eq.$\left( \ref{Moyal}\right) $ involving both particles and field. For
instance we might label the coordinates and momenta of the N particles with
subindices $j\leq 3N$ and the variables $\left\{ x_{j},p_{j}\right\} $
associated to the (classical) field amplitudes $\left\{ a_{j}\right\} $ via
eq.$\left( \ref{10}\right) $ for $j>3N,$ which would allow using eq.$\left( 
\ref{Moyal}\right) $ for both particles and field. In this case the
Hamiltonian eq.$\left( \ref{7}\right) $ could be written 
\[
H_{tot}=H\left( \left\{ x_{k},p_{k}\right\} \right) , 
\]
where now the generalized coordinates and momenta may correspond to either
particles or to radiation modes of the field. The Hamiltonian $H\left(
\left\{ x_{k},p_{k}\right\} \right) $ might be got from eq.$\left( \ref{7}%
\right) $ via a change of variables.

This WW formalism might allow calculating non-relativistic radiative
corrections, like the non-relativistic parts of the Lamb shift or the
anomalous magnetic moment of the electron which ha been calculated also
using the canonical quantization, see e.g. the book of Milonni \cite{Milonni}%
.

\subsection{Discussion}

Here I shall comment on three aspects of the proposed treatment of
non-relativistic quantum electrodynamics within the WW formalism. That is
mathematical, conceptual and practical.

From the \textit{mathematical} point of view the unified treatment of
particles and fields looks nice. The joint states and observables of both,
particles and fields, are represented together by a set of generalized
coordinates and momenta, $\left\{ x_{k},p_{k}\right\} $. However the
handling is not completely symmetrical because for the field the ground
state includes the vacuum ZPF eq.$\left( \ref{ZPF}\right) $, which is not
the case for particles.

For \textit{conceptual }point of view I mean whether the formalism provides
an intuitive description of reality, that is whether the fomalism allows a
realistic interpretation. It is the case that the WW field formalism does
admits a realistic interpretation, but the WW treatment of particles does
not, as was commented in subsection 2.2. Therefore the non-relativistic QED
in WW does not allow a full realistic interpretation whence its interest is
scarce from the conceptual point of view.

From the \textit{practical }point of view, that is the simplicity of the
calculations, the WW representation of the non-relativistic QED does not
offer advantage over the canonical HS formalism, except maybe when the
particle Hamiltonian is at most quadratic in the coordinates and momenta. In
that case the WW treatment becomes actually classical, that is a combination
of Newtonian dynamics with Maxwell-Lorentz electrodynamics. In practice this
would correspond just to a harmonic oscillator or a set of coupled harmonic
oscillators.

\end{document}